\documentclass[usenatbib]{mn2e}
\input epsf

\newcommand{\rd}{{\rm d}}
\newcommand{\beq}{\begin{equation}}
\newcommand{\eeq}{\end{equation}}
\newcommand{\beqary}{\begin{eqnarray}}
\newcommand{\eeqary}{\end{eqnarray}}
\title[The number of Oort Cloud Objects]{Upper limits to the number of Oort Cloud Objects based on serendipitous occultation events search in X-rays} 
\author[Chang, Liu, and Shang]{Hsiang-Kuang Chang$^{1,2}$\thanks{E-mail: hkchang@phys.nthu.edu.tw}, Chih-Yuan Liu$^1$, and Jie-Rou Shang$^1$
\\
$^{1}$Institute of Astronomy, National Tsing Hua University, 
Hsinchu 30013, Taiwan\\ 
$^{2}$Department of Physics, National Tsing Hua University, 
Hsinchu 30013, Taiwan} 

\begin{document}

\date{Accepted 2016 July 19. Received 2016 July 17; in original form 2016 May 30}

\pagerange{\pageref{firstpage}--\pageref{lastpage}} \pubyear{2016}

\maketitle

\label{firstpage}

\begin{abstract}
Using all the RXTE archival data of Sco X-1 and GX 5-1, which amount  to about 1.6 mega seconds in total,
we searched for possible occultation events caused by Oort Cloud Objects. 
The detection efficiency of our searching approach was studied with simulation.
Our search is sensitive to object size of about 300 m in the inner Oort Cloud, 
taking 4000 AU as a representative distance,
and of 900 m in the outer Oort Cloud, taking 36000 AU as the representative distance.
No occultation events were found in the 1.6 Ms data. 
We derived upper limits to the number of Oort Cloud Objects, 
which are about  three orders of magnitude higher than the highest theoretical estimates in the literature for the inner Oort Cloud,
and about six orders higher for the outer Oort Cloud.
Although these upper limits are not constraining enough, they are the first obtained observationally, without
making any model assumptions about comet injection. 
They also provide guidance to such serendipitous occultation event search in the future.    
\end{abstract}


%SSSSSSSSSSSSSSSSSSSSSSSSSSSSSSSSSSSSSSSSSSSSSSSSSSSSSSSSSS
\section{Introduction}
Existence of the Oort Cloud, which was proposed 66 years ago 
\citep{oort50}, has not yet been directly confirmed, 
except for the discovery of Sedna \citep{brown04}, 2012 VP$_{113}$ \citep{trujillo14}
and three other objects \citep{chen13,brasser15}, which are considered to be inner Oort Coloud Objects.
The population properties of Oort Cloud Objects
are not yet clear either, such as their total number and size distribution.
The knowledge of these properties is important to our understanding of
the origin and evolution of the Oort Cloud.

Oort Cloud Objects are usually thought to form in the outer planetary region and
to be ejected to a distance far away from the sun by giant outer planets 
(for brief reviews, see, e.g., \citet{dones04a,rickman14,dones15}).
Galactic tides, star/giant-molecular-cloud encounters, and planetary perturbations
all affect the formation and evolution of the Oort Cloud as well as the injection of Oort Cloud Objects 
into the inner solar system. 
With a proper injection model and observed flux of long-period comets, one can in principle
infer the number of objects in the Oort Cloud.
This has been being the effort made by many authors with various physical considerations.
In \citet{duncan87} it was found, by direct numerical integration, that the Oort Cloud has a sharp inner edge at
about 3000 AU and the
number density of Oort Cloud Objects falls steeply outwards, roughly as $r^{-3.5}$.
The Oort Cloud is therefore centrally condensed, with roughly 5 times more objects
in the inner Oort Cloud ($a < 20,000$ AU) than in the classical outer Oort Cloud.
In a later study, \citet{weissman96} estimated that there are
$10^{12}$ objects of diameter larger than 2.3 km in the outer Oort Cloud ($a > 20,000$ AU).
The inner Oort Cloud was originally proposed by \citet{hills81}.
\citet{kaib09} demonstrated that a significant fraction of long-period comets in fact come from
the inner Oort Cloud before their migration into the outer Oort Cloud due to planetary perturbation.
It was estimated in \citet{kaib09} that there could be $\sim 10^{12}$ objects in the inner Oort Cloud (3000 AU $<a<$ 20,000 AU).
Some estimates yielding smaller numbers of objects in the outer Oort Cloud, about $2\times 10^{11}$ -- $5\times 10^{11}$, were 
reported in \citet{heisler90,dones04b,brasser13}.

Direct observation of objects in the Oort Cloud region is extremely difficult.
Their existence, however, in addition to being inferred from observations of long-period comets, 
may also be explored by occultation events that they cause to distant background stars. 
Such an approach, looking for serendipitous occultation events from a vast amount of data, 
has been being employed to study the size distribution of small Kuiper Belt Objects (KBOs) 
down to sub-kilometer size using optical data \citep{roques06,bickerton08,schlichting12,zhang13,liu15} 
and even to decameter size using X-ray observations \citep{chang13}.
 It has also been applied to estimating upper limits to the number of small objects at distances between 100 and 1000 AU with 
data accumulated by the TAOS project \citep{wang09}.

In this paper we report the result of our effort to extend the work of serendipitous KBO occultation search
in X-rays to the regime relevant to possible occultation events caused by Oort Cloud Objects.
Our conclusion is that in all the RXTE archival data of Sco X-1 and GX 5-1, 1.6 mega seconds in total, 
no such events were found. 
The derived upper limits to the number of inner and outer Oort Cloud Objects
are about three orders of magnitude higher than the highest theoretical estimates in the literature for the inner,
and about six orders higher for the outer, respectively.
Although these upper limits are not constraining enough, they are the first obtained observationally, without
making any model assumptions about comet injection. 
They also provide guidance to observations in the future.
At the end of this paper, we discuss the possibility of using ASTROSAT \citep{singh14}, Athena (a future ESA X-ray space mission \citep{ayre15})
and LOFT (a proposed mission \citep{zane14}) to conduct such a search to study the size distribution of Oort Cloud Objects and KBOs.  
 
%SSSSSSSSSSSSSSSSSSSSSSSSSSSSSSSSSSSSSSSSSSSSSSSSSSSSSSSSS
\section{Searching for occultation events in X-rays caused by small Oort Cloud Objects}

It is clear that small Trans-Neptunian Objects (TNOs, referred to all the solar system objects
beyond Neptune, including KBOs and Oort Cloud Objects)
cannot be observed directly. Roughly speaking, an object of V-band magnitude 30 is
about 10 km in diameter if it is at 50 AU, and about 1000 km at 500 AU
(see e.g. \citet{roques09}). 
Serendipitous occultation search is so far the only way to `observe' objects smaller than that.
Optical search has the advantage of monitoring many stars at the same time, while in X-rays
even smaller objects can be probed because of less diffraction in the occultation light curve.
To date, such search is focused on pinning down the size distribution of small KBOs 
\citep{roques06,bickerton08,schlichting12,chang13,zhang13,liu15}, 
which are located mainly between 30 and 50 AU. 
We now extend this effort to the Oort Cloud, using X-ray data.

To perform such serendipitous occultation search in X-rays, the background X-ray source must be
bright enough or the detector effective area needs to be large enough so that there are
enough detected photons to allow statistically significant determination of occurrence of occultation events and, 
if possible, to allow meaningful determination of some information from fitting the occultation
light curve with diffraction patterns of given parameters. 
The Proportional Counter Array (PCA) on board RXTE has the largest effective area among all the 
astronomical X-ray instruments until the launch of ASTROSAT in 2015 and Sco X-1 is the brightest X-ray source in the sky other than the Sun.
The PCA-observed spectrum of Sco X-1 peaks at 4 keV, for which
the Fresnel scale is 30 m at 40 AU. The shadow of KBOs of that Fresnel-scale size will sweep over the Earth in milliseconds.
The RXTE/PCA count rate of Sco X-1 is rouhly $10^5$ cps, which is about enough for examining possible events
at millisecond time scales. 
That is why in earlier attempts \citep{chang06,chang07,chang11,chang13} only RXTE data of Sco X-1 was employed to 
look for occultation events caused by KBOs down to decameter size. 
Other background sources are all too dim for such millisecond-time-scale search (cf.\ Table 3 in \citet{chang13}).

For Oort Cloud Objects, we take 4000 AU and 36000 AU as two representative distances
for the inner and outer cloud respectively. 
The corresponding Fresnel scales are 300 m and 900 m
and the relevant time scale is then tens of milliseconds. 
We note that, although shadow size decreases with decreasing object size, its diameter
 will asymptotically approach 3.4 Fresnel scales
for small objects \citep{nihei07}, assuming the background source is a point one.
We therefore conducted the search with light curves binned at 10, 30, 60, 90 and 180 ms. 
With such bin sizes, we added another source, GX 5-1, which is the second brightest according to RXTE/ASM count rates 
(cf.\ Table 3 in \citet{chang13}),
to conduct this search. 
Their count rates are of considerable variation, depending on observation conditions and the intrinsic variability of the sources.
We examined all of their RXTE/PCA data, collected data of adequate data modes, and selected data of average count rates in a pointing larger than 4000 cps
for our search. 
The count rate distribution of the data we used is shown in Table \ref{datacps}.
%TTTTTTTTTTTTTTTTTTTTTTTTTTTTTTTTTTTTTTTTTTTTTTTTTTTTTTTTTTTTT
\begin{table}
\begin{center}
\begin{tabular}{crcr}
\hline
\multicolumn{2}{c}{Sco X-1} & \multicolumn{2}{c}{GX 5-1} \\
\hline
Count rates (cps) & Data & Count rates (cps) & Data \\
\hline
4,000 - 20,000 & 50.6 ks & 4,000 - 6,000 &  4.9 ks \\
20,000 - 40,000 & 134.3 ks & 6,000 - 8,000 & 203.1 ks \\
40,000 - 60,000 & 154.7 ks & 8,000 - 10,000 & 115.0 ks \\
60,000 - 80,000 & 282.2 ks & 10,000 - 12,000 & 115.7 ks\\
80,000 - 100,000 & 226.2 ks & 12,000 - 14,000 & 73.9 ks\\
100,000 - 120,000 & 127.8 ks & 14,000 - 16,000 & 72.2 ks\\
120,000 - 140,000 & 13.0 ks & 16,000 - 18,000 & 2.5 ks\\
\hline
Total & 988.9 ks & Total & 587.2 ks  \\
\hline
\end{tabular}
\end{center}
\caption{The count-rate distribution of the RXTE/PCA Sco X-1 and GX 5-1 data employed in this work. Total data volume is 1.6 Ms.
The average count rate for Sco X-1 is about 80000 cps and for GX 5-1 about 8000 cps.}
\label{datacps}
\end{table}
%TTTTTTTTTTTTTTTTTTTTTTTTTTTTTTTTTTTTTTTTTTTTTTTTTTTTTTTTTTTTT

\begin{figure}
\epsfxsize=8.8cm
\epsffile{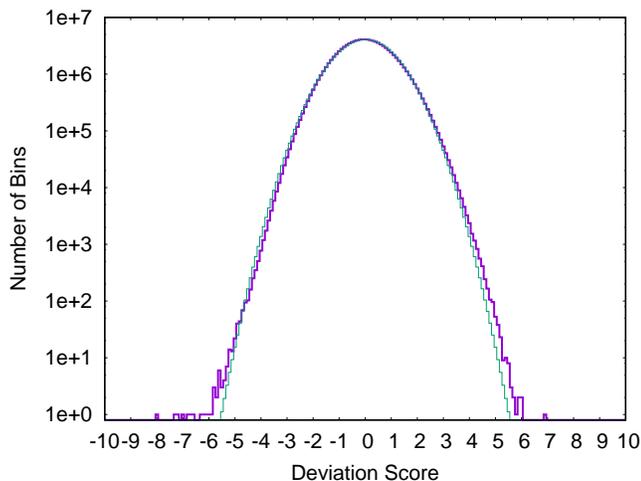}
\caption{The deviation distribution of all the light curves from the 1.6-Ms data binned at 10 ms (thick purple histogram).
A Gaussian distribution is plotted for comparison (thin green histogram). 
The data, because of its Poisson nature, is not expected to follow the Gaussion distribution strictly.
The six events with deviation scores below $-6.5$ are apparently outliers in this distribution. 
Another event with a deviation score of $-11.8$ is not shown here.
More details of these events are discussed in the text and shown in Figure \ref{evtlc}.
The event at $+6.9\sigma$ is statistically significant and will be further studied, but it is not relevant to the subject of this paper.
}
\label{devdis}
\end{figure}

Our searching algorithm is the same as the one we used before (e.g. \citet{chang07,chang11}), 
but with data binned at 
a larger time scale. 
Briefly speaking, we examine their light curves binned at 10, 30, 60, 90 and 180 ms
to look for bins with statistically significant photon-count drop.
We assign to each bin a deviation score, which is the difference between the photon count number of that bin and the average count number per bin 
in an 8-sec running window in units of the standard deviation in that 8-sec window.
To judge whether a dip event is consistent with random fluctation, we compare the deviation distribution of all bins in a search at a certain bin size
with a Gaussian distribution. 
In all of our search, we found seven bins with deviation stronger than $-6.5\sigma$. 
All of them are found in the case of 10-ms bin size, whose deviation distribution is  shown in Figure \ref{devdis}. 
These seven bins are apparently outliers in that deviation distribution. 
The level at $-6.5\sigma$ corresponds roughly to a random probability of $10^{-3}$ if a Gaussion distribution is assumed and the total number of bins
is taken into account.
%FFFFFFFFFFFFFFFFFFFFFFFFFFFFFFFFFFFFFFFFFFFFFFFFFFFFFFF
%FFFFFFFFFFFFFFFFFFFFFFFFFFFFFFFFFFFFFFFFFFFFFFFFFFFFFFFF 

\begin{figure*}
\epsfxsize=7.6cm
\epsffile{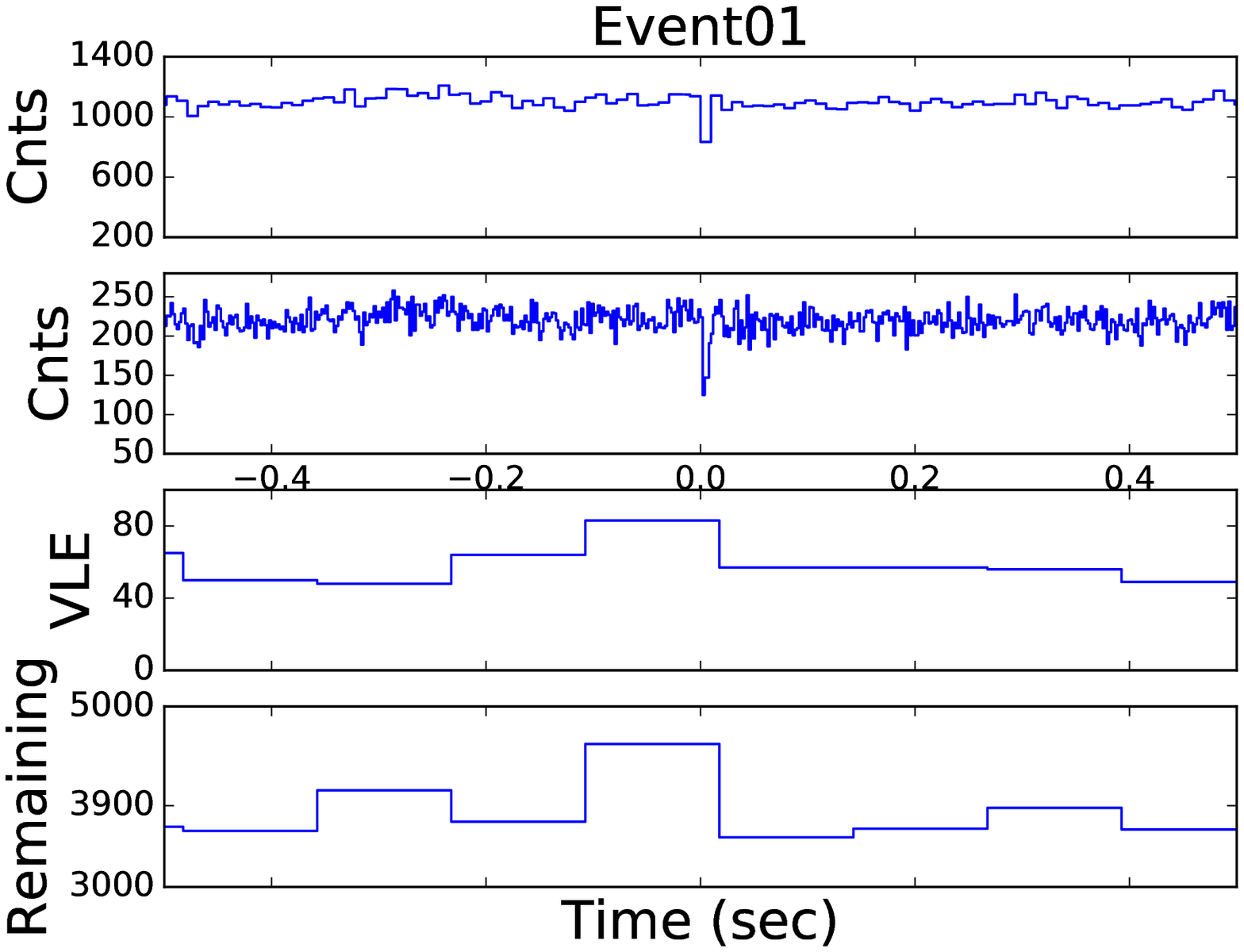}
\epsfxsize=7.6cm
\epsffile{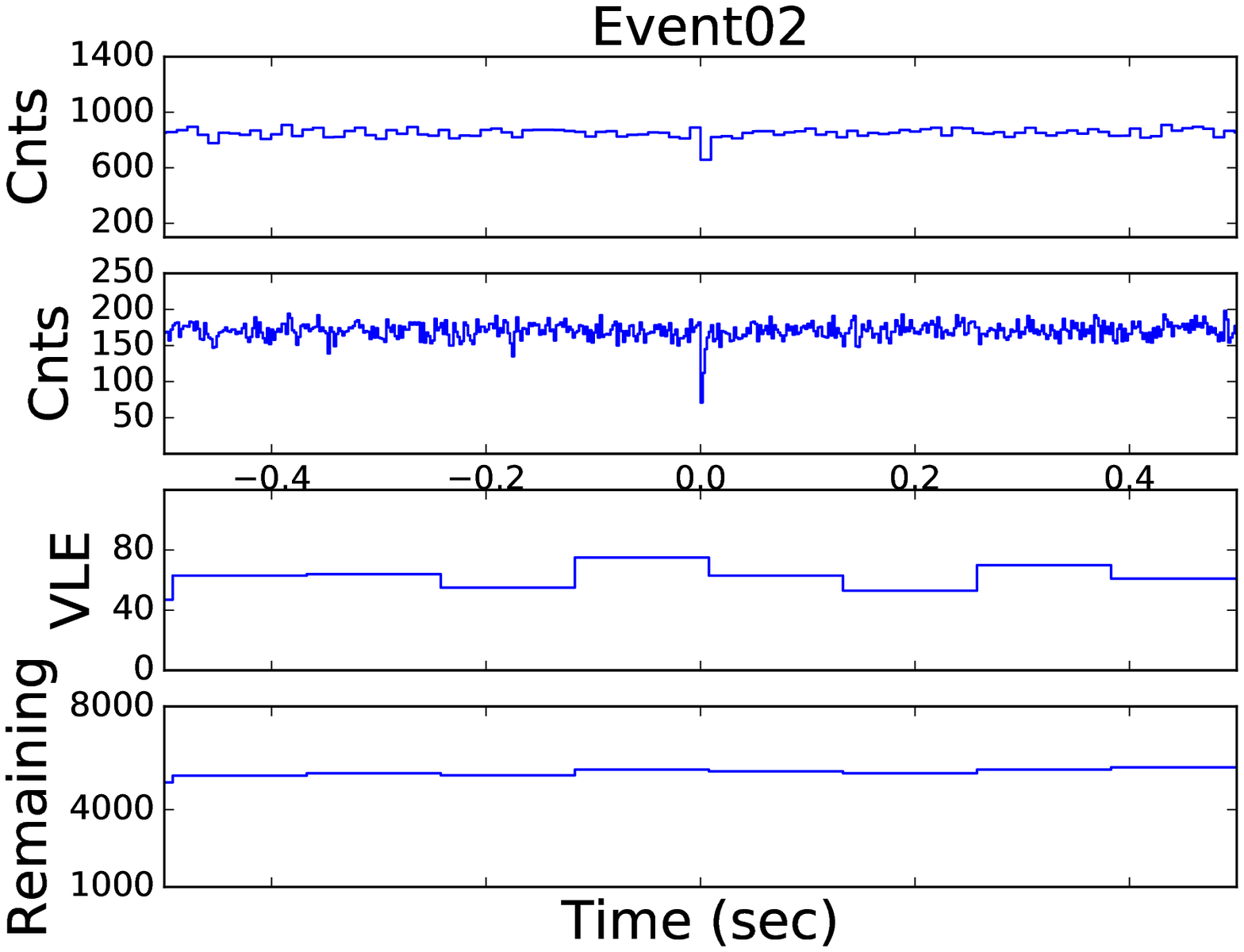}
\epsfxsize=7.6cm
\epsffile{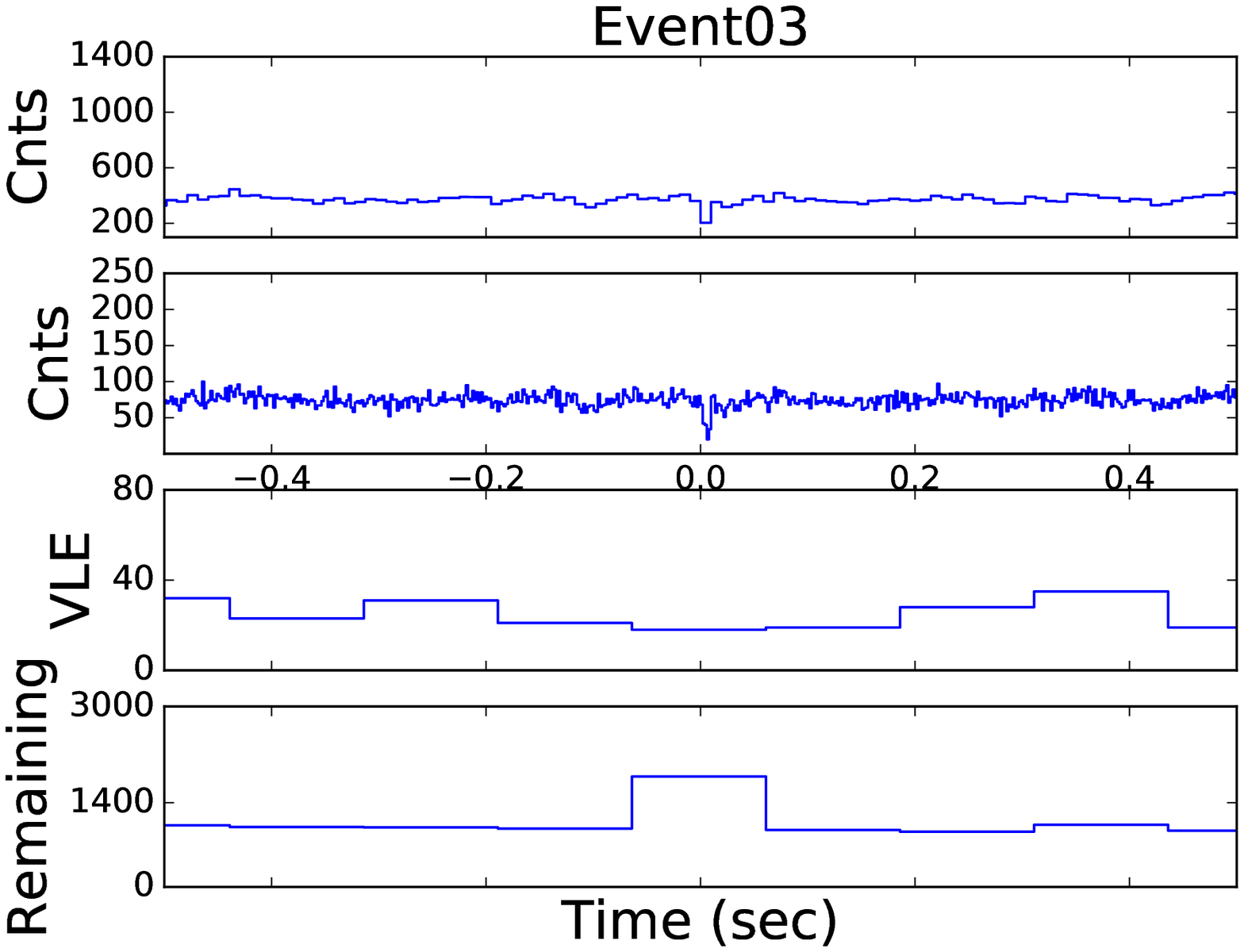}
\epsfxsize=7.6cm
\epsffile{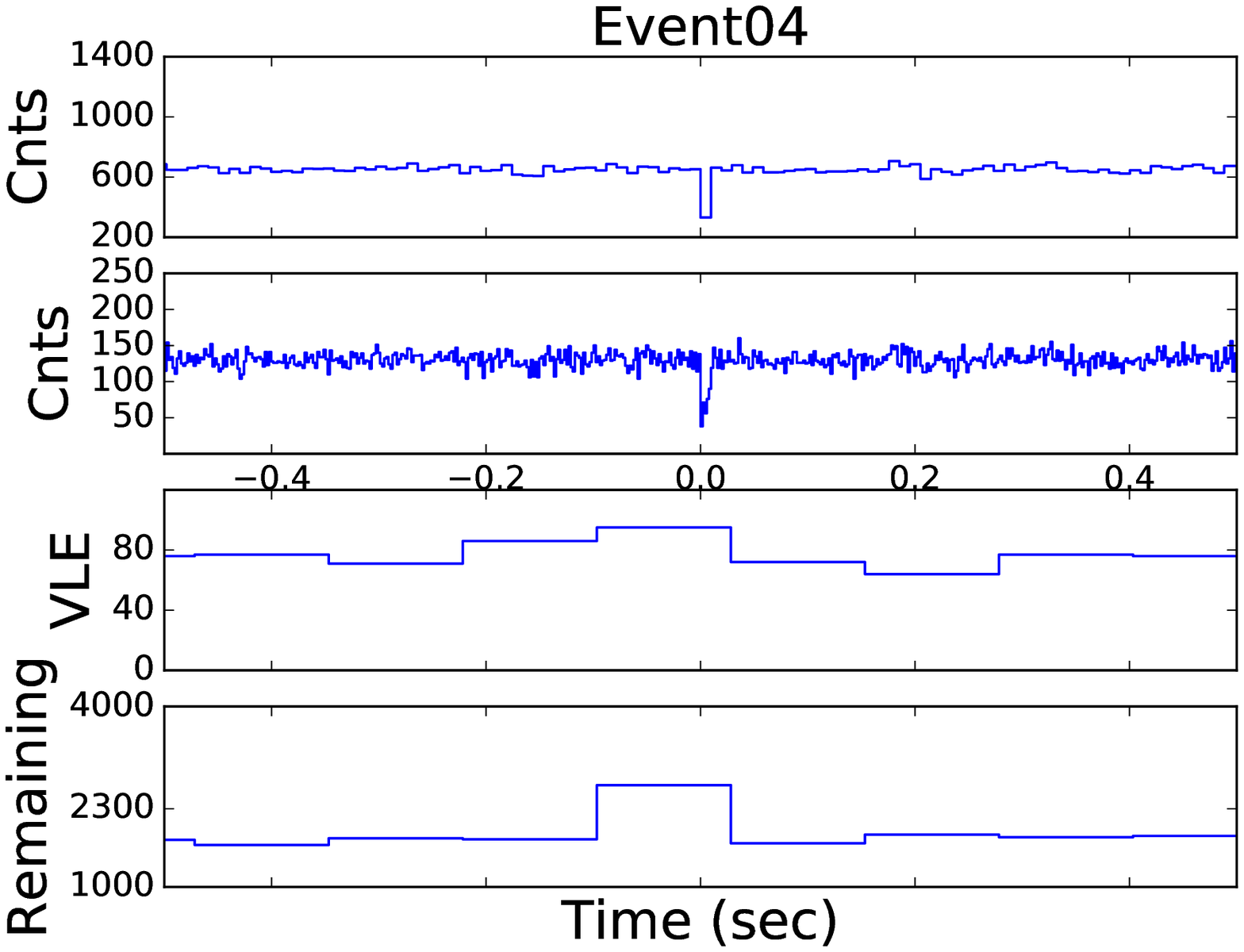}
\epsfxsize=7.6cm
\epsffile{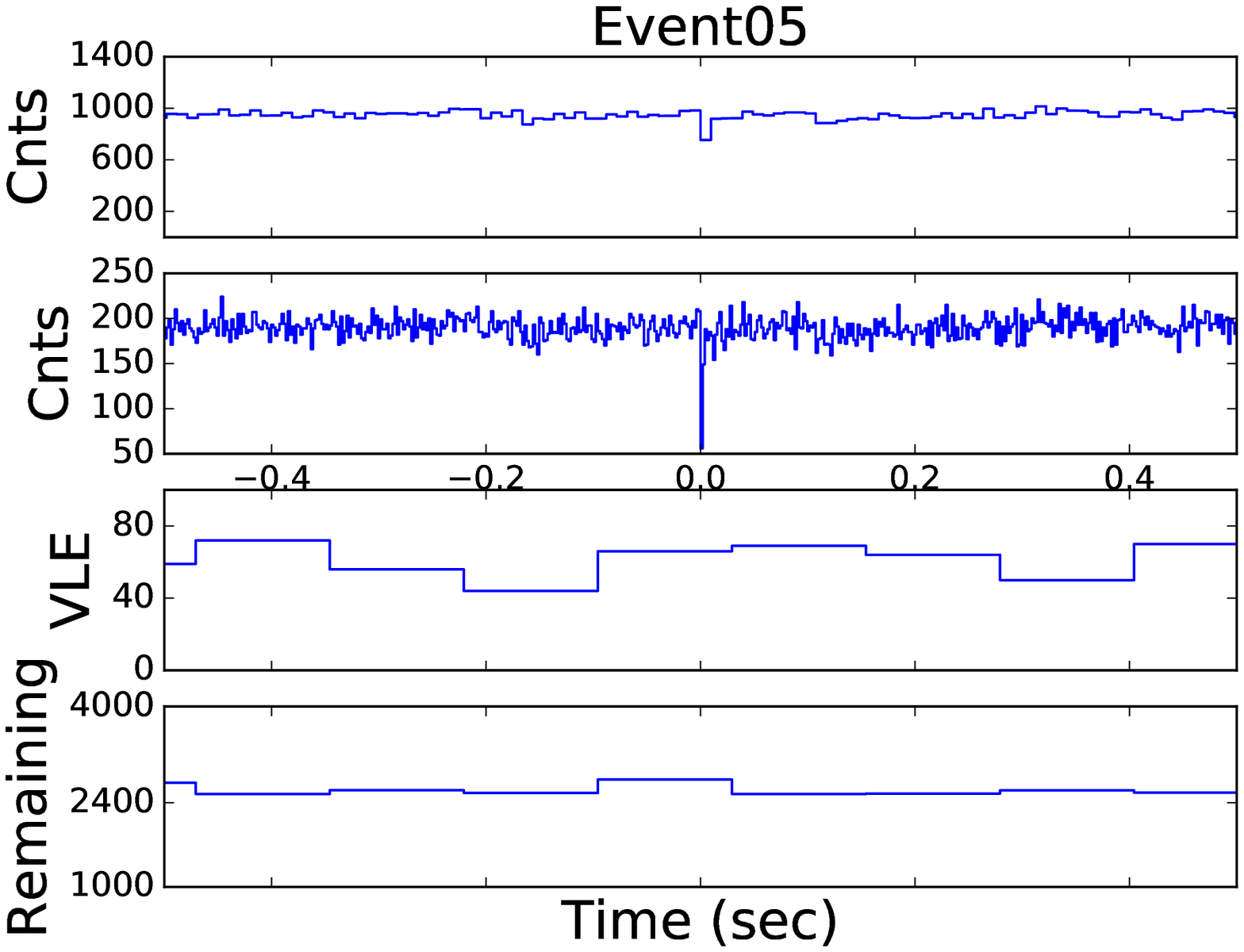}
\epsfxsize=7.6cm
\epsffile{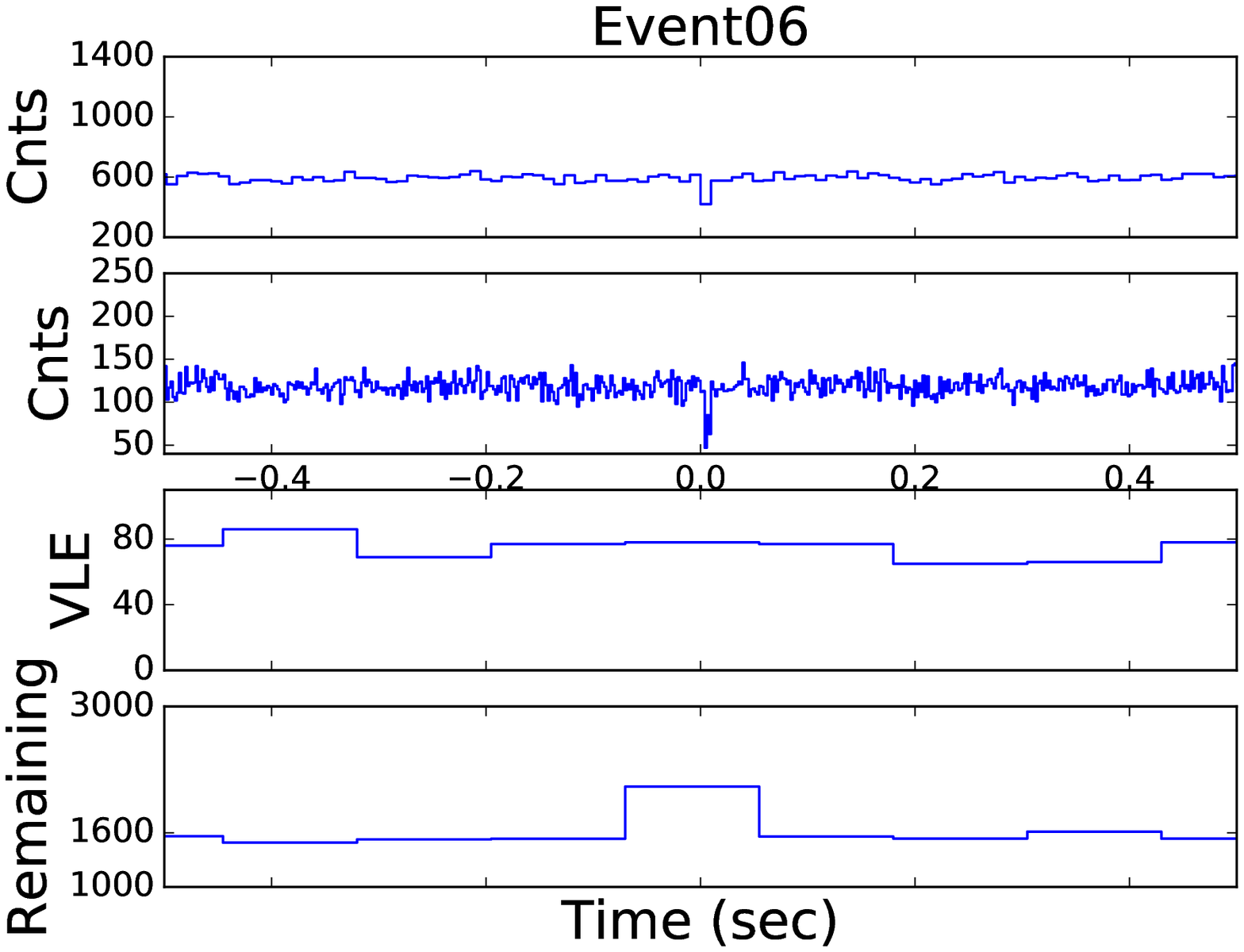}
\epsfxsize=7.6cm
\epsffile{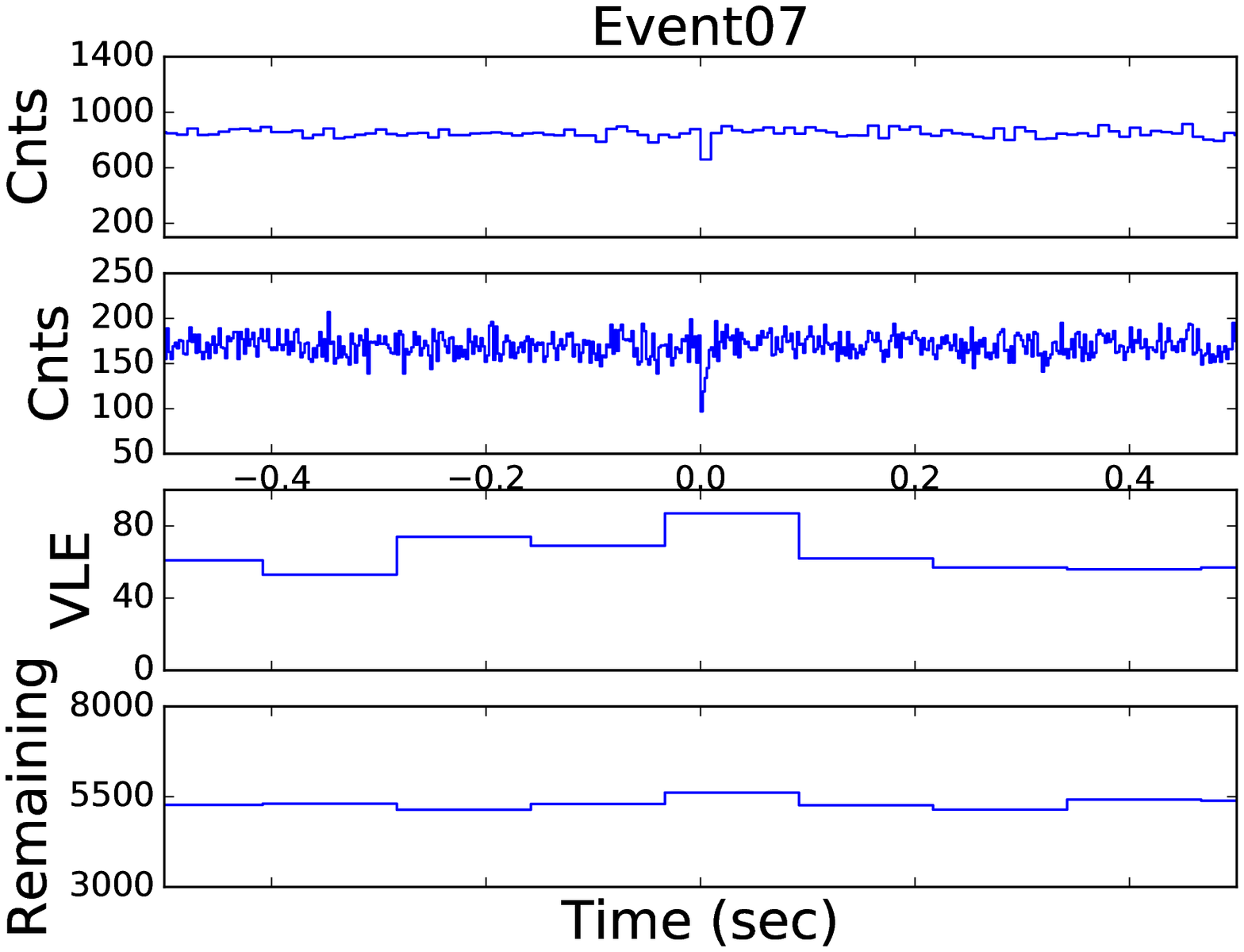}
\caption{Light curves of the seven dip events in a 1-s window. Shown here for each event, from top to bottom, are the light curves binned at 10 ms and 2 ms, 
and of the `VLE' and `Remaining' counts, which are only available with 125-ms resolution in the RXTE/PCA housekeeping data. See the text for the explanation
of `VLE' and `Remaining' counts. 
}
\label{evtlc}
\end{figure*}

More details of these seven events are shown in Table \ref{7evt} and Figure \ref{evtlc}.
%TTTTTTTTTTTTTTTTTTTTTTTTTTTTTTTTTTTTTTTTTTTTTTTTTTTTTTTTTTTTT
\begin{table}
\begin{center}
\begin{tabular}{cccc}
\hline
Event & Epoch (MJD) & RXTE ObsId & Significance \\
\hline
1 & 50228.94786301 & 10056-01-02-00 & $-7.3\sigma$  \\
2 & 50820.70967736 & 30036-01-01-000 & $-8.0\sigma$   \\
3 & 52413.98833994 & 70015-01-01-01 & $-7.4\sigma$  \\
4 & 54333.91603852 & 93067-01-08-00 & $-11.8\sigma$  \\
5 & 54765.73618319 & 93067-01-28-00 & $-6.7\sigma$  \\
6 & 55699.14184696 & 93067-04-20-00 & $-6.6\sigma$ \\
7 & 55928.58787662 & 96443-03-01-05 & $-7.0\sigma$  \\
\hline
\end{tabular}
\end{center}
\caption{Epochs of the seven events found in the 10-ms light curves.}
\label{7evt}
\end{table}
%TTTTTTTTTTTTTTTTTTTTTTTTTTTTTTTTTTTTTTTTTTTTTTTTTTTTTTTTTTTTT
%FFFFFFFFFFFFFFFFFFFFFFFFFFFFFFFFFFFFFFFFFFFFFFFFFFFFFFF
%FFFFFFFFFFFFFFFFFFFFFFFFFFFFFFFFFFFFFFFFFFFFFFFFFFFFFFFF 
Among these seven events, Events 4, 5 and 6 are associated with events found at smaller bin size reported in \citet{chang13} that come together
with the so-called Very Large Events (VLEs). 
RXTE's VLEs are events that
deposit more than 100 keV into an individual anode in a Proportional
Counter Unit (PCU). RXTE/PCA consists of five identical
PCUs. An instrument dead-time of 50 $\mu$s is set for each VLE during
RXTE observations of Sco X-1. VLEs are produced by high energy
photons or particles. In order to clarify whether VLEs cause the dip events reported in \citet{chang06,chang07}, 
a special data mode was designed for the RXTE observation proposal 93067 to record each individual VLE with $\mu$s resolution and information 
of all the triggerred anodes. In usual housekeeping data, VLEs are recorded only with its total number in a 125-ms bin.
It was found that all the millisecond dip events, except for one, are associted with VLEs. 
They are considered instrumental, although not expected at the beginning with a 50-$\mu$s dead time.
As can be seen in Figure \ref{evtlc}, these three events all have an asymmetric profile in their 2-ms light curves.
They also all have
some excess in the `Remaining' counts, which are all the events except for ‘good’ events, VLEs and
propane events. Good events are those that trigger only one xenon
anode. The propane event is the one that triggers only the propane layer, which is designed to distinguish events caused
by soft electrons from those by X-rays. The ‘remaining’ counts
contain many ‘multiple-anode’ events, which are usually treated as
particle background. Excess in the `Remaining' counts is indication of instrumental effects.
The other four events, that is, Events 1, 2, 3, and 7 are all either with asymmetric profiles in thier 2-ms light curves 
or with some 'Remaining' count excess.
Besides, all these seven events are found in Sco X-1's light curves. None is in GX 5-1's. 
These events all show a deep drop in the light curve, but not down to zero, in a certain 2-ms bin, and then gradually 'recover' back to normal.
Such a dip, when binned at 10 ms, has only a modest equivalent depth. For GX 5-1, with a typical count number of 80 in a 10-ms bin, 
such a dip will not have a deviation score as large as -6.5, because of its small count number in one bin.  
We therefore conclude that they are all instrumental.
No dip events were found in the 1.6 Ms data with light curves binned at 10-, 30-, 60-, 90- and 180-ms.

%FFFFFFFFFFFFFFFFFFFFFFFFFFFFFFFFFFFFFFFFFFFFFFFFFFFFFFF
\begin{figure*}
\epsfxsize=16.0cm
\epsffile{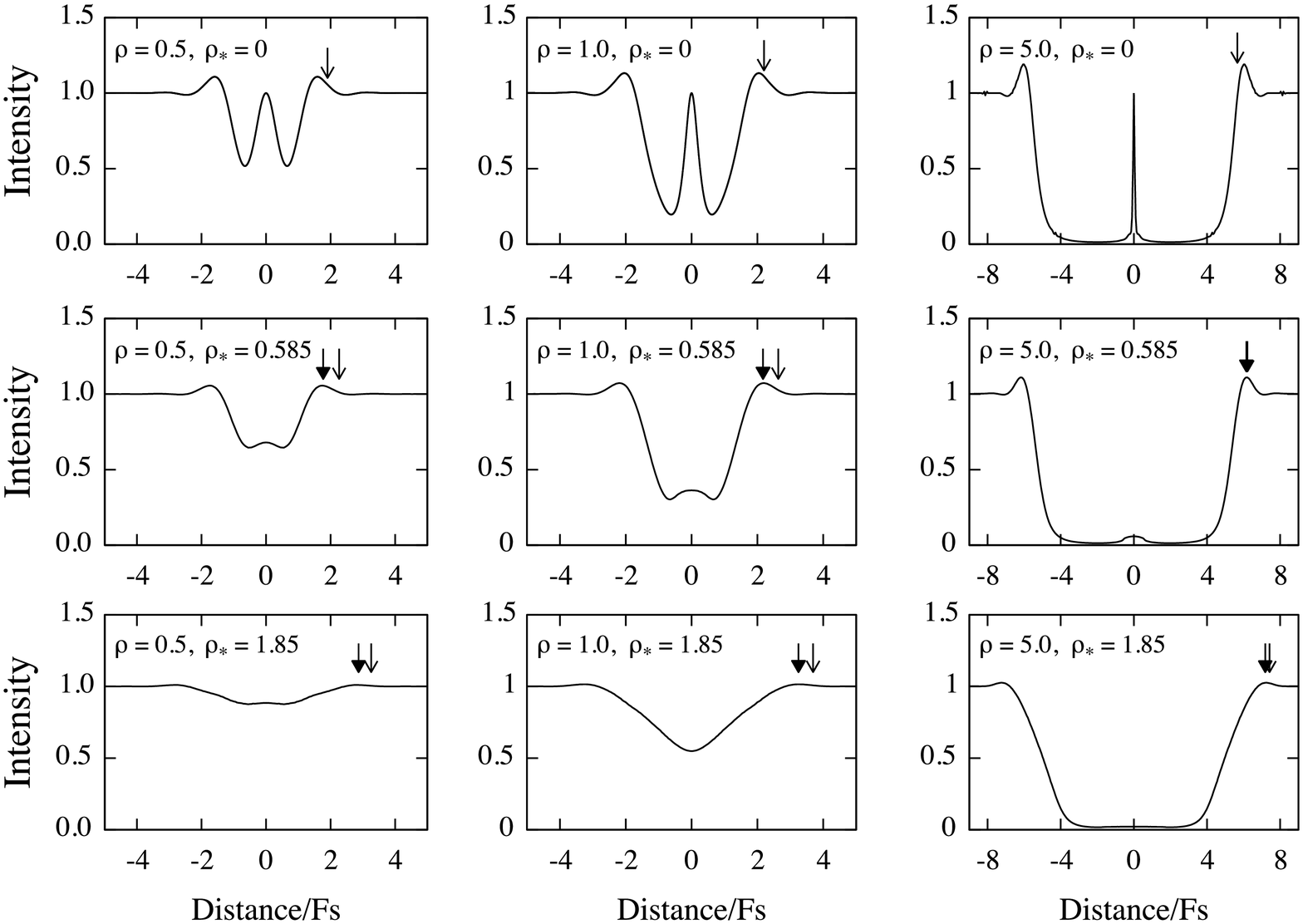}
\caption{Some example diffraction patterns of central-crossing events for different sizes of the occulting body and the background source.
Shown here are that for occulting body radius 0.5 Fs (panels in the left column), 1.0 Fs (the middle column) and 5.0 Fs (the right column).
The panels in the top row are for background sources being a point source, in the middle row for radius 0.585 Fs, and in the bottom row radius 1.85 Fs.
Thick arrow symbols indicate the shadow boundary we adopt, which is at the fring peak. 
Thin arrow symbols indicate the shadow boundary as given by Eq.(4) in \citet{nihei07} for extended background sources.
In the middle right panel, the two arrows almost coincident.
In the top row, for a point-like background source ($\rho_*=0$), the thin arrows indicate the shadow boundary as 
described by $\Omega=2(\sqrt{3}^{3/2}+\rho^{3/2})^{2/3}$ \citep{nihei07}. We adopt this in our simulation for the case of
point-like background sources. 
A typical RXTE/PCA observed Sco X-1 spectrum is used for the computation. 
}
\label{pattern}
\end{figure*}
%FFFFFFFFFFFFFFFFFFFFFFFFFFFFFFFFFFFFFFFFFFFFFFFFFFFFFFFF 

%SSSSSSSSSSSSSSSSSSSSSSSSSSSSSSSSSSSSSSSSSSSSSSSSSSSSSSSSS
\section{RXTE detection efficiency for occultations caused by small Oort Cloud Objects}

%FFFFFFFFFFFFFFFFFFFFFFFFFFFFFFFFFFFFFFFFFFFFFFFFFFFFFFF
\begin{figure}
\epsfxsize=8.4cm
\epsffile{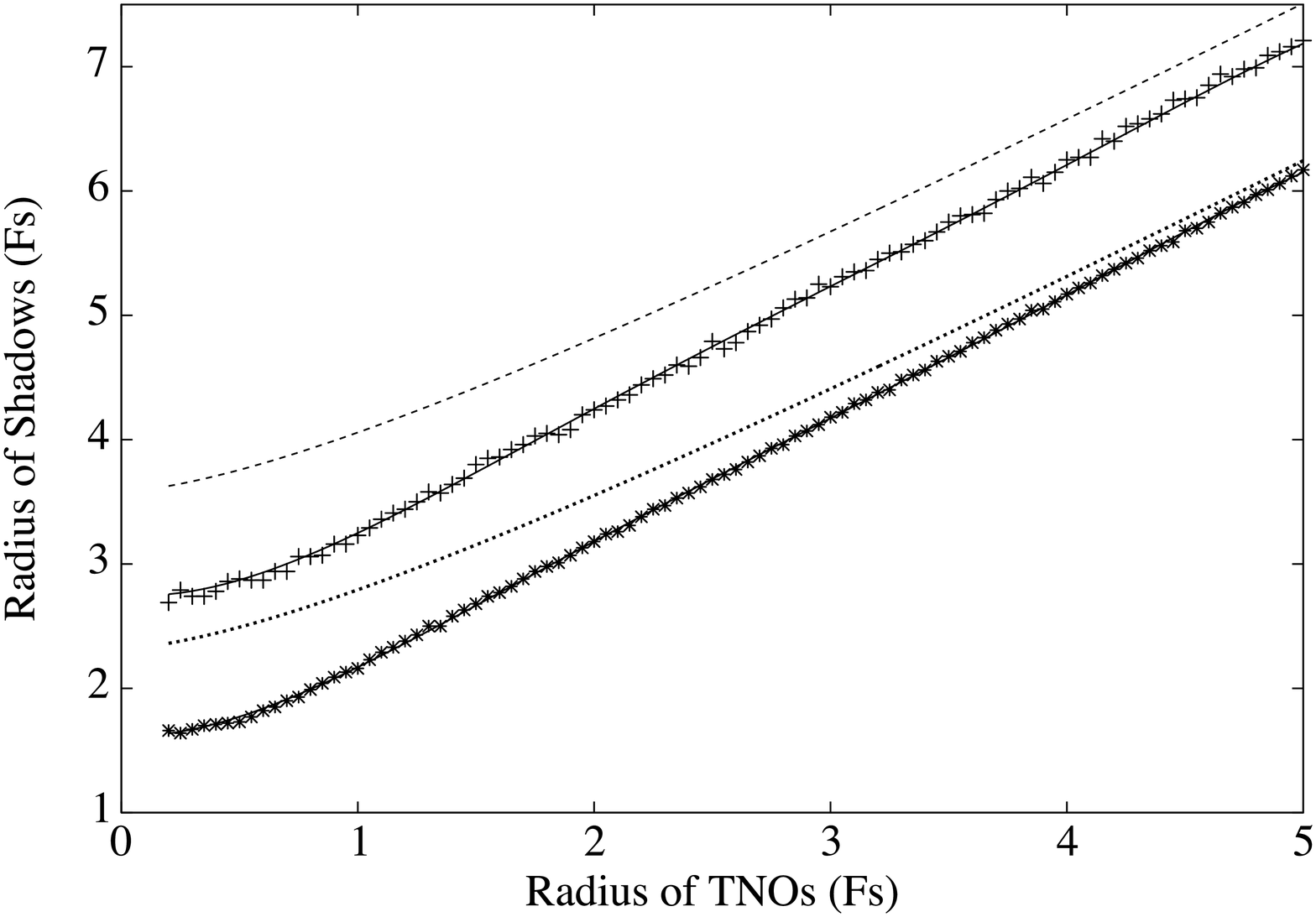}
\caption{The shadow radius as a function of TNO radius for extended background sources.
Shown with asterisk symbols (the lower curve)  is the case for the extended source radius being 0.585 Fs, and with cross symbols (the upper curve) for radius 1.85 Fs.
A polynomial fit, which is used in our detection efficiency simulation, is shown with solid lines passing through those symbols.
Dotted and Dashed lines are those obtained from Eq.(4) in \citet{nihei07} for the cases of 0.585 Fs and 1.85 Fs, respectively.
}
\label{shara}
\end{figure}
%FFFFFFFFFFFFFFFFFFFFFFFFFFFFFFFFFFFFFFFFFFFFFFFFFFFFFFFF 
To derive the size distribution from detected event rates, 
we need to know the detection efficiency of the searching approach.
In \citet{chang13}, it was studied for the case of detecting KBO occultation events 
with RXTE data of Sco X-1. In this paper we do the same but for the case of detecting
Oort Cloud Object occultation events, which involves different time scales in the 
searching algorithm and different shadow diffraction patterns because of the
possibly non-negligible size of the X-ray emitting region in Sco X-1 and GX 5-1 as projected at the
distance of the Oort Cloud. 

The detection efficiency, $\eta$, depends on the `shadow size' that one adopts, 
detector count rate of the background source, relative speed, and the searching algorithm.
In the geometric optics regime the shadow size is simply the diameter $D$ of the occulting body,
if the background star can be well approximated as a point source.
If the size of the background star is not negligible, the full shadow size, 
including umbra and penumbra, will be
$D+D_*$, where $D_*$ is the projected diameter of the background star 
at the distance of the occulting body.
When diffraction matters, it is larger and not well defined.
For a point-like background source, taking the fringe peak as the shadow boundary, 
it was found that
\citep{nihei07}
the relation between
the shadow size $\Omega$ and the occulting body radius $\rho$ can be approximatedly
described as $\Omega=2(\sqrt{3}^{3/2}+\rho^{3/2})^{2/3}$, 
in which
both $\Omega$ and $\rho$ are those in units of Fresnel scale.  
For extended background sources, it was suggested to be 
$\Omega=2(\sqrt{3}^{3/2}+\rho^{3/2})^{2/3}+2\rho_*$ (Eq.(4) in \citet{nihei07}), where $\rho_*$ is the projected radius of the
background star at the distance of the occulting body.

The size of the X-ray emitting region in Low Mass X-ray Binaries, such as Sco X-1 and GX 5-1, is not yet well determined.
It could range from about 50 km to 50000 km, depending on different models
(e.g., see \citet{barnard03,bradshaw03,revnivtsev14,hynes16} for Sco X-1, and \citet{jackson09,sriram12} for GX 5-1).
For the largest end, 50000 km, the projected size  
at 40, 4000 and 36000 AU for Sco X-1, whose distance is 2.8 kpc, is roughly 3.6, 360 and 3240 m respectively, which is about
0.12, 1.2, and 3.6 Fresnel scale. 
The Fresnel scale (Fs hereafter) is defined as $\sqrt{\lambda d/2}$, where $\lambda$
is the wavelength and $d$ is the distance. 
It is about 30 m for $d=40$ AU and $\lambda=0.3$ nm 
(4 keV; most photons in the RXTE data of Sco X-1 and GX 5-1 are at about this energy).
The distance to GX 5-1 is not quite certain. From column density estimate, it is about 9 kpc \citep{christian97}.
In the Oort Cloud region, therefore, we should consider the possibility of 
an extended background source. 

We computed shadow patterns of Fresnel-scale sized TNOs for three background-target
sizes: a point source, 0.585 Fs radius, and 1.85 Fs radius, with a typical 
RXTE-observed Sco X-1 spectrum. Example shadow patterns of TNOs 
with 0.5 Fs, 1.0 Fs, and 5.0 Fs radius
are shown in Figure \ref{pattern}, in which the shadow-size range we adopt is indicated,
which uses the diffraction fringe peak as the boundary of the shadow.
The numerically computed shadow size as a function of TNO size is plotted
in Figure \ref{shara}. This shadow size is what we use in our simulation for
determining detection efficiency. In practice we use a polynomial fit to describe it.
We note that the expression in \citet{nihei07} (Eq.(4)) gives a shadow size
larger than that defined by the diffraction fringe peaks, 
particularly for smaller occulting bodies. It is also shown in Figure \ref{shara}.

When estimating the detection efficiency with simulations, as detailed below, 
we randomly pick an impact parameter $\beta$, 
ranging from zero to the shadow radius, in each simulation.
The impact parameter $\beta$ is the distance between the center 
of the background star and that of the occulting body.
In principle one may consider a larger range for the impact parameter, 
but in that case the detection efficiency
will be lower and more efforts are required to obtain a statistically stable solution. 
What really matters in deriving the size distribution (see Eq.(\ref{evrt})
 is the product of the detection efficiency and the shadow size, the latter of which 
is represented by the impact-parameter
range in the simulation. 
In this sense, the shadow size expression for extended background sources as suggested in \citet{nihei07} may also
be used. 
The product $\eta\Omega$ should have the same value for different choices of $\Omega$, 
which defines 
the impact-parameter range, if they are all chosen to be large enough. 

%FFFFFFFFFFFFFFFFFFFFFFFFFFFFFFFFFFFFFFFFFFFFFFFFFFFFFFF
\begin{figure}
\epsfxsize=8.4cm
\epsffile{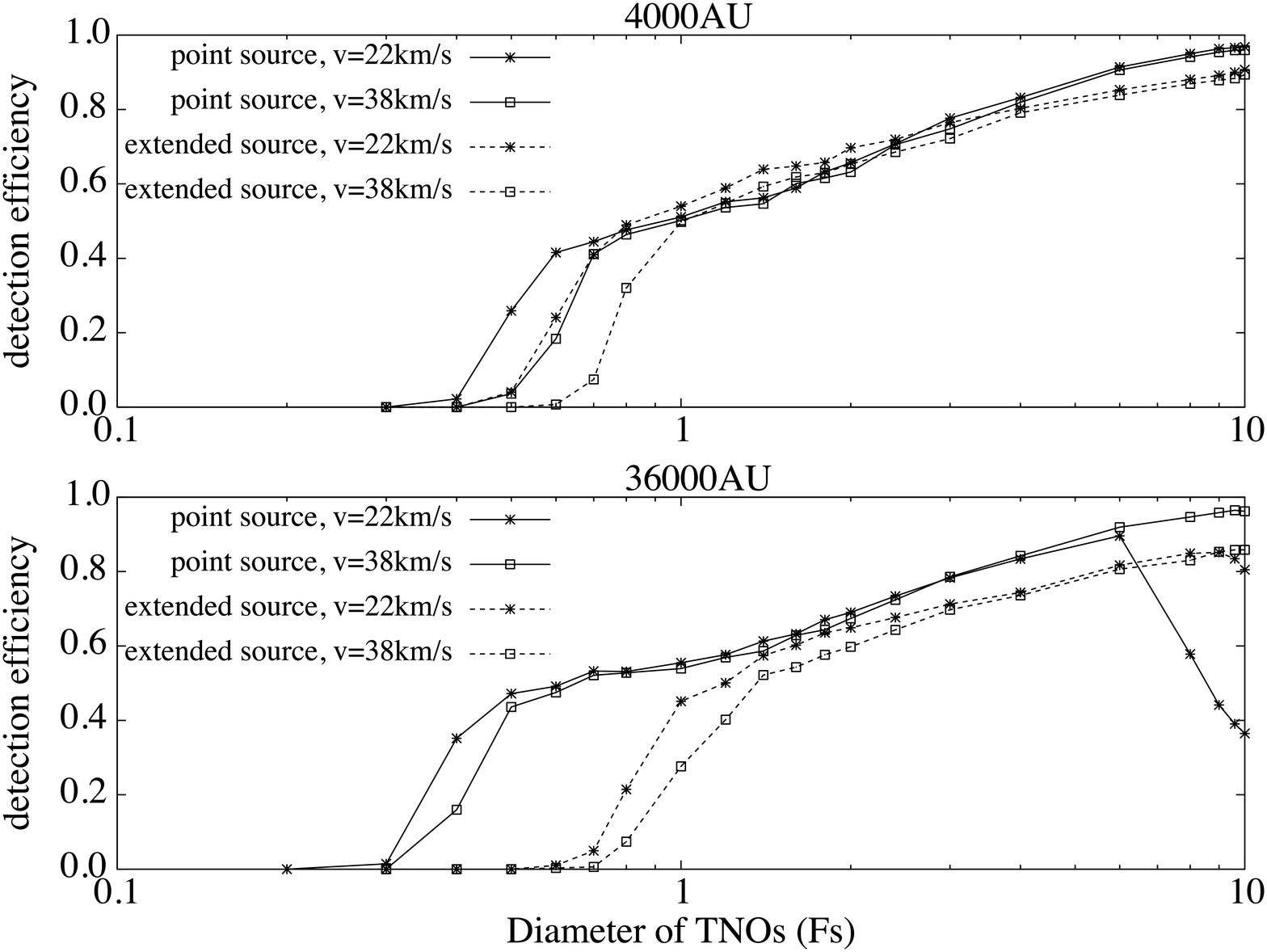}
\caption{The detection efficiency as a function of TNO diameter. 
Solid lines are for the background source being a point one and dashed lines for extended background sources.
In the upper panel the diameter of the extended background source is 1.2 Fs (360 m), corresponding to 50000 km at 2.8 kpc projected at 4000 AU,
and in the lower panel it is 3.6 Fs (3240 m), corresponding to that at 36000 AU. 
The lines with asterisk symbols are for the relative speed being 22 km/s and those with open squares for 38 km/s.
The drop of the solid line with asterisk symbols close to 10 Fs in the lower panel is discussed in the text. 
}
\label{etascox1}
\end{figure}
%FFFFFFFFFFFFFFFFFFFFFFFFFFFFFFFFFFFFFFFFFFFFFFFFFFFFFFFF 
%FFFFFFFFFFFFFFFFFFFFFFFFFFFFFFFFFFFFFFFFFFFFFFFFFFFFFFF
\begin{figure}
\epsfxsize=8.4cm
\epsffile{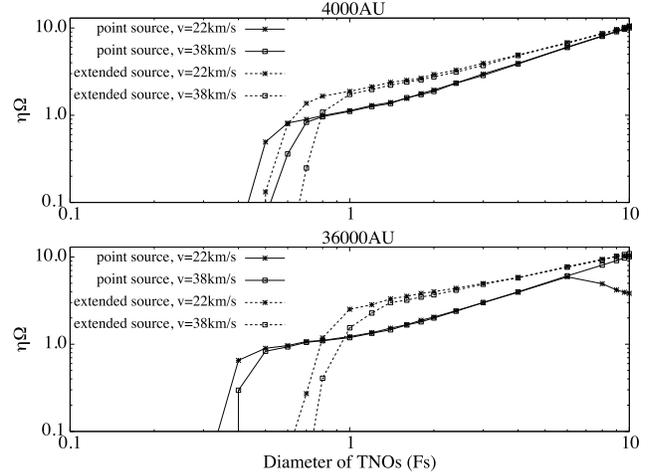}
\caption{The product of detection efficiency and shadow diameter as a function of TNO diameter.  Figure legend is the same as that in
Figure \ref{etascox1}. The shadow diameter is the one we adopt as shown in Figure \ref{shara}.
}
\label{etaomegascox1}
\end{figure}
%FFFFFFFFFFFFFFFFFFFFFFFFFFFFFFFFFFFFFFFFFFFFFFFFFFFFFFFF 
In the following we present simulation results of detection efficiency 
for cases of assuming
the occulting body being at 4000 AU (inner Oort Cloud) and 36000 AU (outer Oort Cloud).
We used typical RXTE/PCA-observed spectra of Sco X-1 and GX 5-1, 
which both peak at about 4 keV, 
as the input spectra when computing the diffraction pattern. 
The count rate is set at 80000 cps for Sco X-1 and at 8000 cps for GX 5-1.
In both the cases of inner and outer Oort Cloud, 
we further distinguish the possibility of treating
the background sources as a point-like source or as an extended source.
As discussed above, the X-ray emitting region in these sources is not well determined. It is model dependent and could range from about 50 km to 50000 km.
Sco X-1 is at 2.8 kpc away. A scale of 50 km at 2.8 kpc corresponds to 0.36 m at 4000 AU and 3.2 m at 36000 AU. 
The Fresnel scale is 300 m at 4000 AU and 900 m at 36000 AU. It is therefore still a good approximation to treat
Sco X-1 as a point-like source if its X-ray emitting region is close to 50 km. The same also applies to GX 5-1, whose distance is likely about 9 kpc.
On the other hand, if it is 3 orders larger, that is, close to 50000 km, it will be 1.2 Fresnel scale at 4000 AU and
3.6 Fresnel scales at 36000 AU for Sco X-1. For Gx 5-1 it would be three times smaller if we still assume a 50000-km X-ray emitting region.
It is, however, not really meaningful to distinguish the size of the possible extended regions between Sco X-1 and GX 5-1, because that is very uncertain. 
We therefore use those values for the cases of 
`extended sources' in our simulation for both Sco X-1 and GX 5-1.

In each simulation with a given relative speed and the occulting body size, 
we performed 10000 runs of event search in the
simulated light curve. In each run, we randomly pick the impact parameter 
$\beta$ in the range from zero to the shadow radius
(see discussion at the beginning of this section) 
and the epoch of the implanted occultation event.
We conducted the same algorithm as we used in searching for dip events with real data binned at different bin size.
The threshold of detection is set at $-6.5\sigma$.
Figure \ref{etascox1} shows the results of
the detection efficiency for the case of Sco X-1. 
 The actuall relative speed between the occulting body and the RXTE satellite for each possible event
depends on the velocity of RXTE in the solar system at the event epoch and that of the occulting body. For very distant objects, the latter can be neglected. 
Since RXTE observations of Sco X-1 were conducted more intensively when Sco X-1 is close to opposition and the RXTE orbital speed relative to the Earth is about 7.8 km/s, we consider relative speeds being
$30\pm 8$ km/s in our simulation. Exact relative speed can be found for specific candidate occultation events \citep{chang11}.
There are some points to note:
First, roughly speaking, the detection efficiency
gets larger than 50\% for occulting TNO size (diameter) larger than one Fresnel scale
for all cases.
Second, even for TNOs as large as of 10-Fs diameter, the detection efficiency 
is still noticeably below 100\%. This is because of those cases
with impact parameters close to the shadow radius. It is more apparent when
the background target is extended, since the corresponding diffraction pattern
has a more gradual transition boundary, and therefore there is a shallower shadow for a larger range of large impact parameters; 
see the right panels of Figure \ref{pattern} and compare the boundary transition from the top panel to the bottom.
Third, in the lower panel of Figure \ref{etascox1}, we see an obvious drop of the detection
efficiency for the point-source lower-speed case close to 10-Fs diameter. 
This is because the shadow duration, which is about 450 ms for a central-crossing 
event when the relative speed is 22 km/s,
is much larger than the bin size used in the search, the largest of which is 180 ms.
The shadow causes an overestimate of the standard deviation in the search running window
and suppresses the chance of detection.     
For such cases, the more gradual transition boundary caused by an extended 
background target actually reduces that overestimate and the suppression is
therefore weaker, as shown for the lower-speed extended-source case in the lower panel of Figure \ref{etascox1}.
Fourth, the results for GX 5-1 are similar to that of Sco X-1, except that the 50\% detection efficiency 
occurs at an TNO diameter of about 2 Fs, rather than 1 Fs. This is of course due to the lower count rate.
      
As discussed earlier and also shown in Eq.(\ref{evrt}), 
when estimating the event rate, what matters is the product 
$\eta\Omega$ as a function of TNO diameter $D$,
which is plotted in Figure \ref{etaomegascox1}.
One can see that $\eta\Omega$ is close to $D$/Fs for $D$ larger than about 1 Fs. 
Based on all these results, we consider one Fresnel scale as
an appropriate smallest size that our approach can probe.
In the next section we further approximate $\eta\Omega$ with $D$/Fs for
$D$ larger than 1 Fs. This is somewhat underestimated, and therefore the
inferred upper limit is overestimated a little bit.

%SSSSSSSSSSSSSSSSSSSSSSSSSSSSSSSSSSSSSSSSSSSSSSSSSSSSSSSSS
\section{Upper limits to the number of Oort Cloud Objects}

To describe the size distribution of a certain population, such as MBAs, KBOs or Oort Cloud Objects,
one often uses the accumulated number  of objects larger than a certain size per unit sky area,
$n_{D>D_1}$, defined as
\beq
n_{D>D_1}=\int^{D_1}_\infty\frac{\rd n}{\rd D}\,\rd D
\,\,\,\, ,
\eeq
where $\frac{\rd n}{\rd D}$ is the differential size distribution.
In collisional equilibrium, one expects to have \citep{obrien03} 
$\frac{\rd n}{\rd D}\propto D^{-q}$
and $q = \frac{7 + p/3}{2 + p/3}$,
where p is the power index of the strength-size relation. 
In the so-called ‘strength-scaled regime’ (in contrast to the ‘gravity-scaled
regime’ for larger bodies), p is negative.
It has been found that the size distribution of MBAs has a wavy shape,
or, approximately, the power index $q$ takes different values in different size ranges
(see, e.g., \citet{yoshida07}). 
Similar behavior was also tentatively found for KBOs (see, e.g., \citet{liu15}).
Within a certain size range in which a constant power index $q$ is a good approximation,
the accumulated number density and the differential number density is related as
\beqary
\lefteqn{n_{D_2>D>D_1}  =} \nonumber \\
%& & \left(\frac{\rd n}{\rd D}\right)_{D_0} D_0\frac{1}{-q+1}
%((\frac{D_1}{D_0})^{-q+1}-(\frac{D_2}{D_0})^{-q+1})\nonumber \\
   &  &\left(\frac{\rd n}{\rd \log D}\right)_{D_0} \frac{1}{\ln 10}\frac{1}{-q+1}((\frac{D_1}{D_0})^{-q+1}-(\frac{D_2}{D_0})^{-q+1}), 
\eeqary
where the definition
\beq
\frac{\rd n}{\rd D}=\left(\frac{\rd n}{\rd D}\right)_{D_0}\left(\frac{D}{D_0}\right)^{-q}
\eeq
is adopted.

The event rate of serendipitous occultation caused by TNOs of diameter $D$ in the range
of $D_2>D>D_1$, for one background source, can be estimated as
\beq
\frac{N}{T}=\int^{D_1}_{D_2}\frac{\rd n}{\rd D}\frac{(\eta\Omega\,\mbox{Fs})\, v}{d^2}\,\rd D\,\left(\frac{180}{\pi}\right)^2
\,\,\,\, ,
\label{evrt}
\eeq
where $N$ is the number of events, $T$ the total observing time, $n$ in units of number per square degree,
$\eta\Omega$ is the product of detection efficiency and the shadow size as described in the previous section,
$v$ the relative speed between TNO and the observer,
and $d$ the distance.
In this paper we approximate $\eta\Omega$ with $D$/Fs 
for $D$ larger than one Fresnel scale; more details are described in
the previous section, the study of detection efficiency.
We do not have detailed information about $v$ and $d$ as functions of $D$.
Typical, or representative, values are assumed to estimate the event rate.

Since we do not have any knowledge about the size distribution of Oort Cloud Objects,
let us first assume a constant power index $q$ for objects with diameter larger than 
one Fresnel scale, denoted as $D_1$ below, which is clsoe to the smallest
size to which our search is sensible. 
The event rate is then
\beqary
\lefteqn{\frac{N}{T}=\left(\frac{\rd n}{\rd \log D}\right)_{D_0}\frac{1}{\ln 10}} \nonumber \\
 & & \frac{1}{-q+2}((\frac{D_1}{D_0})^{-q+2}-(\frac{D_2}{D_0})^{-q+2})\frac{D_0 v}{d^2}(\frac{180}{\pi})^2
\,\, .
\eeqary
Considering $D_2\gg D_1$ and $q>2$, we then have, taking $D_0=D_1$,
\beq
n_{D>D_1}=\frac{N}{T}\left(\frac{-q+2}{-q+1}\right)\frac{d^2}{D_1 v}(\frac{\pi}{180})^2
\,\,\,\, .
\eeq
Since there is no detection in the 1.6 Ms RXTE data that we employed, 
we estimate the upper limit to the size distribution at  95\% confidence level with  $N=3$.
 The Fresnel scale  
at 4000 AU and 36000 AU, for 4 keV photons, is  about 300 m and 900 m respectively.
Taking $v=30$ km/s and $q=3.5$, we then have, for the inner Oort Cloud
\beq
n_{D>300m}< 1.4\times 10^{13}\, \mbox{deg}^{-2}
\,\,\,\, ,
\eeq
and for the outer Oort Cloud 
\beq
n_{D>900m}< 3.6\times 10^{14}\,  \mbox{deg}^{-2}
\,\,\,\, .
\eeq
Assuming an isotropic distribution, the total number is
\beq
N_{D>300m}<5.4\times 10^{17}  \,\,\,\, \mbox{for the inner Oort Cloud} \,\,\,\, ,
\eeq
and
\beq
N_{D>900m}<1.5\times 10^{19}  \,\,\,\, \mbox{for the outer Oort Cloud} \,\,\,\, .
\eeq
These upper limits are way above earlier theoretical estimates in the literature. 
If we consider the estimate  that there are about
$10^{12}$ comets 
of diameter larger than 2.3 km in the outer Oort Cloud \citep{weissman96}
and assume a power index of $q=3.5$, we will have 
$N_{D>900m}=1.1\times 10^{13}$ for the outer Oort Cloud.
This number is about  six orders lower than our upper limit estimate.
If we consider there are five times more objects in the inner Oort Cloud, we have
$N_{D>300m}=8.2\times 10^{14}$ for the inner Oort Cloud.
This number is about three orders smaller than the upper limit we currently have.
Although our upper limits are not stringent enough, they are the first obtained 
by observations. They also provide guidance for future observations of this kind.

%TTTTTTTTTTTTTTTTTTTTTTTTTTTTTTTTTTTTTTTTTTTTTTTTTTTTTTTTTTTTT
\begin{table*}
\begin{center}
\begin{tabular}{lcccccl}
\hline
Sources & \multicolumn{2}{c}{ASTROSAT/LAXPC} & \multicolumn{2}{c}{Athena/WFI}
 & LOFT/LAD  & Other names   \\
%\cline{2-6}
&  (3-12 keV) &  (12-80 keV)
&  (0.2-2 keV)  & (2-12 keV)  
&  (2-12 keV) 
& \\ 
\hline

4U 1758-25 
& $1.5\times 10^4$ &  --- 
& $9.9\times 10^2$  & $1.5\times 10^4$ 
& $2.3\times 10^5$ 
& 1H1758-250, GX5-1   \\

4U 1758-20
& $9.9\times 10^3$  & $1.4\times 10^3$ 
& $2.7\times 10^3$ & $7.0\times 10^3$ 
& $1.5\times 10^5$ 
&  1H1758-205, GX9+1, Sgr X-3 \\

4U 1617-15 
& $1.5\times 10^5$  &  $1.2\times 10^4$  
& $5.3\times 10^5$ & $2.6\times 10^5$ 
& $3.0\times 10^6$        
& 1H1617-155, Sco X-1 \\ 

4U 1813-14
& $7.4\times 10^3$  & ---
& $1.6\times 10^3$  & $7.7\times 10^3$ 
& $1.2\times 10^5$  
& 1H1813-140, GX17+2, Ser X-2 \\ 

4U 1702-36 
& $9.5\times 10^3$  & ---
& $4.5\times 10^3$  & $1.0\times 10^4$ 
& $1.5\times 10^5$  
& 1H1702-363, GX 349+2, Sco X-2 \\ 
 
4U 1642-45 
& $5.3\times 10^3$  & --- 
& $1.8\times 10^2$ & $4.6\times 10^3$ 
& $7.8\times 10^4$ 
& 1H1642-455, GX340+0 \\

4U 1837+04
& $2.8\times 10^3$  & --- 
& $6.1\times 10^3$  & $3.5\times 10^3$
& $4.9\times 10^4$ 
& 1H1837+049, Ser X-1 \\ 

4U 2142+38
& $5.8\times 10^3$  & ---
& $4.5\times 10^3$ & $8.0\times 10^3$
& $1.0\times 10^5$ 
& 1H2142+380, Cyg X-2 \\

4U 1956+35
& $7.3\times 10^3$  & $3.8\times 10^3$ 
& $1.5\times 10^4$  & $8.0\times 10^3$
& $1.2\times 10^5$
& 1H1956+350, Cyg X-1 \\
\hline
\end{tabular}
\end{center}
\caption{
Estimated count rates, in units of count per second (cps),
 for some non-extended bright X-ray sources, 
which are potential 
background targets for serendipitous TNO occultation
search in X-rays. All the nine listed sources are brighter 
than 0.1 Crab in all the 4th Uhuru (4U), HEAO 1 A-1 (A1),
and RXTE All Sky Monitor (ASMquick)
catalogues. The 4U catalog was compiled with observations 
conducted during 1972-1973, the A1 catalog was during 1977, 
and the ASMquick information is based on weekly average in August 2011. 
The estimated count rates are based on spectral models of each sources
reported in the literature (\citet{chang13}, and references therein).
 LAXPC is the 
Large Area X-ray Proportional Counter on board ASTROSAT. 
WFI is the Wide Field Imager on board Athena.
LAD is the Large Area Detector on board LOFT, whose count rate in this table is somewhat different from that in 
\citet{chang13} because  a more updated LOFT/LAD instrument response matrix
(version `M4') is used.
For ASTROSAT/LAXPC, `---' indicates that the source count rate is below three times estimated background count rate.
This list is sorted with the absolute value of their ecliptic latitude.
}
\label{countrate}
\end{table*}
%TTTTTTTTTTTTTTTTTTTTTTTTTTTTTTTTTTTTTTTTTTTTTTTTTTTTTTTTTT

%SSSSSSSSSSSSSSSSSSSSSSSSSSSSSSSSSSSSSSSSSSSSSSSSSSSSSSSSS
\section{Search for occultations in X-rays caused by small TNOs in the future}

To explore  population properties of small bodies in the outskirt of our solar system, 
serendipitous occultation event search seems to be the only way.
Count rates and data volume are the two major factors to consider. 

For KBOs, since the Fresnel scale is about 30 m at 40 AU for 4-keV photons,
one should look for occultation events at millisecond time scale.
For a cetral-crossing event, if taking the relative speed between the observer and the TNO
to be 30 km/s, the shortest shadow duration will be about 3 ms -- recall that diffraction makes the shadow to be
about 3.4 Fs for small occulting bodies. Search with 1-ms time bins is suggested for such a case.
A count rate of about $4\times 10^4$ cps will give 40 counts in a 1-ms bin,
which is about enough for detecting occultation events.
One may of course look for events caused by larger bodies with 
background sources of
lower count rates, but that will probably require much larger data volume 
to have detection or to have
meaningful upper limits to the size distribution, 
since the size distribution is generally believed to decrease
quite fast towards the larger-size end.
We estimated count rates of three X-ray space missions, ASTROSAT, Athena, and LOFT,
 for the nine compact X-ray sources, using detector response matrices
provided by the instrumentation teams and typical source spectra reported in the literature.
One can see that, in Table \ref{countrate}, 
Sco X-1 is still the only usable background source for all the three missions, 
GX5-1, Sco X-2 and Cyg X-1 are marginal ones for Athena, GX 5-1 is also marginal for ASTROSAT, and
all the 9 sources can be used for LOFT.
In our earlier efforts \citep{chang13}, the obtained upper limit with several hundred kiloseconds data
may be already quite close to the real distribution. 
We suggest that a few megaseconds data of these three missions with high enough count rates 
are likely to have some definite detections of such events.

As for Oort Cloud Objects, due to diffraction limit, 
one can only explore objects of sub-kilometer and kilometer
size for inner and outer Oort Cloud Objects respectively. 
For this size range, light curves binned at 10 milliseconds 
are fine enough for occultation event search. 
It therefore only requires a lower count rate of about $4\times 10^3$ cps 
to have 40 counts in one 10-ms bin. 
This why RXTE/PCA data of GX 5-1 can be used in this search 
but not in that for KBOs.
All the nine sources can be used for ASTROSAT, Athena and LOFT for such search, 
except that Ser X-1 may be too dim for ASTROSAT. 
In this paper, however, we report upper limits, with about 1.6-Ms data, 
to the size distribution of inner and outer Oort Cloud Objects,
which are three and six orders higher than the highest theoretical estimates in the literature.
It does not seem feasible for these X-ray space missions 
to accumulate enough data volume to have definite detections, unless
all the earlier theoretical works severely underestimate 
the number of Oort Cloud Objects.

On the other hand,  it is relatively easier to accumulate a huge data volume
in optical bands, see, 
e.g., the Miosotys \citep{shih10}, TAOS II \citep{lehner14} and CHIMERA \citep{harding16} projects.
Because of longer wavelength, however, it can probe objects 
only down to size of 1 km, 10 km and 30 km or so 
for the Kuiper Belt, inner and outer Oort Cloud, respectively,
again due to diffraction limit. 
It in turn requires even more data, say, 
about 300 times more compared with that in X-rays,
depending on the actual size distribution.
 A major limiting factor for ground-based optical occultation survey is the finite angular sizes of the target stars, 
which are typically much larger than occulting bodies and the Fresnel scale in the Oort Cloud region. 
It does not help to have a smaller Fresnel scale by observing at shorter optical wavelength.
Some detections have been reported with fewer data of high signal-to-noise ratio
obtained by space missions like HST \citep{schlichting12} and CoRoT \citep{liu15}. 
These high-quality data make possible the detection of occultation events caused by 
KBOs of about 500-m size, roughly half a Fresnel scale.
It is probably much more difficult to detect occultation events by Oort Cloud Objects.
A dedicated space mission with high quality data
might be able to provide solutions.

%SSSSSSSSSSSSSSSSSSSSSSSSSSSSSS
\section*{Acknowledgments}
Comments from Matthew Lehner to improve this paper are very much appreciated.
This research has made use of data obtained through 
the High Energy Astrophysics Science Archive Research Center Online Service, 
provided by the NASA/Goddard Space Flight Center. 
This work was supported by the Ministry of Science and Technology of 
the Republic of China under grants 101-2923-M-007 -001 -MY3 and 102-2112-M-007 -023 -MY3.

\label{lastpage}
\end{document}